\documentclass[twocolumn,showpacs,floats,floatfix,superscriptaddress,aps,pra]{revtex4}
\usepackage{eurosym}
\usepackage{amsfonts}
\usepackage{amssymb}
\usepackage{amsmath}
\usepackage{graphicx}
\usepackage{bm}
\usepackage{color}
\usepackage{epsfig}
\usepackage{calc}
\usepackage{ifthen}
\usepackage{blindtext}

\begin{document}

\title{Broadband composite polarization rotator}
\date{\today }

\begin{abstract}
We describe a broadband optical device that is capable of rotating the
polarization plane of a linearly polarized light at any desired angle over a
wide range of wavelengths. The device is composed of a sequence of half-wave
plates rotated at specific angles with respect to their fast-polarization
axes. This design draws on an analogy with composite pulses, which is a
well-known control technique from quantum physics. We derive the solutions
for the rotation angles of the single half-wave plates depending on the
desired polarization rotation angle. We show that the broadband polarization
rotator is robust against variations of the parameters of both the crystal
and the light field.
\end{abstract}

\author{Andon A. Rangelov}
\email{rangelov@phys.uni-sofia.bg}
\affiliation{Department of Physics, Sofia University, James Bourchier 5 blvd., 1164
Sofia, Bulgaria}
\affiliation{Engineering Product Development, Singapore University of Technology and
Design, 20 Dover Drive, 138682 Singapore, Singapore}
\author{Elica Kyoseva}
\affiliation{Engineering Product Development, Singapore University of Technology and
Design, 20 Dover Drive, 138682 Singapore, Singapore}
\maketitle



\section{Introduction}


 One of the fundamental properties of light is its polarization \cite{Hecht,Wolf,Azzam,Goldstein,Duarte}, and the ability to observe and manipulate the polarization state is highly desirable for practical applications. For example, there are many optical measurement techniques based on polarization which are used in stress measurements, ellipsometry, physics, chemistry, biology, astronomy and others \cite{Pye,Damask,Landolfi}. Furthermore, the controlled rotation of the light polarization is the underlying principle on which display and telecommunications technologies are based \cite{Matioli}.

Key optical elements for polarization state manipulation are optical polarization rotators, which rotate the light polarization plane at a desired angle \cite{Hecht,Wolf,Azzam,Goldstein,Duarte}. The commercially available polarization rotators, which typically exploit the effect of circular birefringence, possess two main advantages. First, the angle of light polarization rotation is independent of rotation of the rotator around its own optical axis; and second, they are fairly inexpensive. However, they are useful only for a limited range of wavelengths. Alternatively, polarization plane rotation can be achieved through consecutive reflections, which is the underlying principle of Fresnel rhombs \cite{Hecht,Wolf,Azzam,Goldstein,Duarte}. They posses the advantages of operating over a wide range of wavelengths but are quite expensive.

A scheme to rotate the light polarization plane for several fixed
wavelengths was suggested by Koester \cite{Koester}. The design presented
in Ref. \cite{Koester} uses up to four half-wave plates in a series to ensure good rotation of the polarization plane for four different
wavelengths. Another possible method to enlarge the spectral width of a
light polarization plane rotation is to use two wave plates of the same
material as shown by Kim and Chang in Ref. \cite{Kim}.

In this paper, we further extend the approach of Kim and Chang \cite{Kim} in
combination with the Koester idea \cite{Koester} and design an arbitrary broadband
polarization rotator which outperforms existing rotators for broadband operation. Our scheme
consists of two crossed half-wave plates, where the angle between their fast
axes is half the angle of polarization rotation. We design the two half-wave
plates to be broadband using composite pulses approach \cite%
{Ivanov,Peters,Rangelov,Kyoseva}. That is, each composite half-wave plate
consists of a number of individual wave plates arranged at specific angles
with respect to their fast axes. This ensures that the proposed polarization
rotator is broadband and stable with respect to wavelength variations. We
provide the recipe for constructing a polarization rotator device such that, in principal, an arbitrary broadband polarization profile can be achieved.


\section{Background}


Any reversible polarization transformation can be represented as a
composition of a retarder and a rotator \cite{Hurvitz}. A rotation at an
angle $\theta $ is represented by the Jones matrix in the
horizontal-vertical (HV) basis as,
\begin{equation}
\mathfrak{R}(\theta )=\left[
\begin{array}{cc}
\cos \theta & \sin \theta \\
-\sin \theta & \cos \theta%
\end{array}%
\right] .
\end{equation}
A retarder is expressed in the HV basis by the Jones matrix,
\begin{equation}
\mathfrak{J}(\varphi )=\left[
\begin{array}{cc}
e^{i\varphi /2} & 0 \\
0 & e^{-i\varphi /2}%
\end{array}%
\right] ,
\end{equation}%
where the phase shift is $\varphi =2\pi L(n_{\mathnormal{f}}-n_{\mathnormal{s%
}})/\lambda $, with $\lambda $ being the vacuum wavelength, $n_{\mathnormal{f%
}}$ and $n_{\mathnormal{s}}$ the refractive indices along the fast and slow
axes, correspondingly, and $L$ the thickness of the retarder. The most widely-used
retarders are the half-wave plate ($\varphi =\pi $) and the quarter-wave
plate ($\varphi =\pi /2$). The performance of such retarders is usually
limited to a narrow range of wavelengths around $\lambda$ due to their
strong sensitivity to variations in the thickness and the rotary power of
the plate.

Let us now consider a single polarizing birefringent plate of phase shift $%
\varphi $ and let us present a system of HV polarization axes (HV basis),
which are rotated at an angle $\theta $ with reference to the slow and the
fast axes of the plate. The Jones matrix $\mathfrak{J}$ then has the form
\begin{equation}
\mathfrak{J}_{\theta }(\varphi )=\mathfrak{R}(-\theta )\mathfrak{J}(\varphi )%
\mathfrak{R}(\theta ).  \label{jones2}
\end{equation}
In the left-right circular polarization (LR) basis this matrix attains the
form $\mathbf{J}_{\theta }(\varphi )=\mathbf{W}^{-1}\mathfrak{J}%
_{\theta}(\varphi )\mathbf{W}$, where $\mathbf{W}$ connects the HV and LR
polarization bases,
\begin{equation}
\mathbf{W}=\tfrac{1}{\sqrt{2}}\left[
\begin{array}{cc}
1 & 1 \\
-i & i%
\end{array}
\right] .
\end{equation}
Explicitly, the Jones matrix for a retarder with a phase shift $\varphi $
and rotated at an angle $\theta $ is given as (in the LR basis),
\begin{equation}
\mathbf{J}_{\theta }(\varphi )=\left[
\begin{array}{cc}
\cos \left( \varphi /2\right) & i\sin \left( \varphi /2\right) e^{2i\theta }
\\
i\sin \left( \varphi /2\right) e^{-2i\theta } & \cos \left( \varphi /2\right)%
\end{array}
\right] .  \label{retarder}
\end{equation}
For example, half- and quarter- wave plates rotated at an angle $\theta $, $%
(\lambda /2)_{\theta }$ and $(\lambda /4)_{\theta }$, are described by $%
\mathbf{J}_{\theta }(\pi )$ and $\mathbf{J}_{\theta }(\pi /2)$, respectively.


\section{Composite broadband half-wave plate}

\label{sec3}

\begin{figure}[htb]
\centerline{\includegraphics[width=0.95\columnwidth]{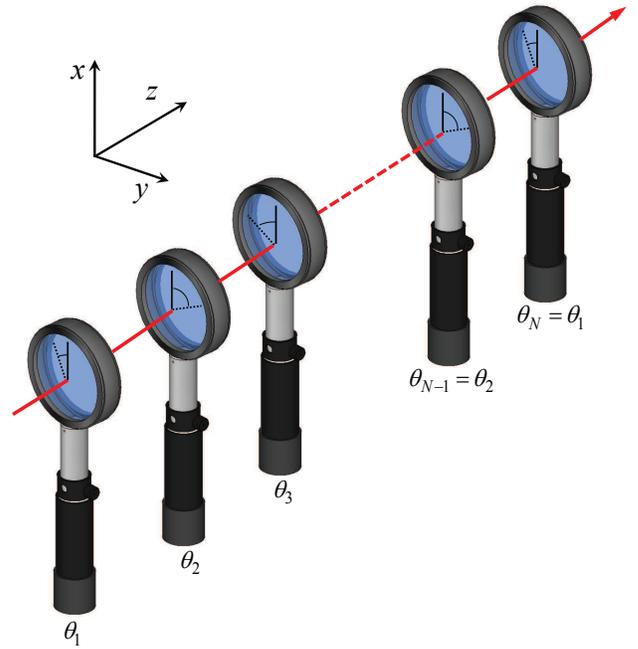}}
\caption{(Color online) Schematic structure of the composite broadband
half-wave plate, which consist of a stack of $N$ ordinary half-wave plates
rotated at specific angles $\protect\theta _{k}$. The fast polarization axes
of the wave plates are represented by dashed lines, while the solid lines
represent the $x$ direction of the coordinate system.}
\label{Fig1}
\end{figure}

Our first step is to design a half-wave plate that is robust to variations
in the phase shift $\varphi $ at $\varphi =m\pi $ ($m=1,2,3...$). Such
half-wave plates allow for imperfect rotary power $\varphi /L$ and
deviations in the plate thickness $L$, and furthermore, operate over a wide
range of wavelengths $\lambda $. To achieve this, we will follow an
analogous approach to that of composite pulses \cite%
{Ivanov,Peters,Rangelov,Kyoseva}, which is widely adopted for robust control
in quantum physics \cite{Torosov1,Genov,Torosov2}. In detail, we replace the
single half-wave plate with an arrangement of an odd number $N=2n+1$
half-wave plates (shown schematically in Fig.\ref{Fig1}). Each wave plate
has a phase shift $\varphi =\pi $ and is rotated at an angle $\theta _{k}$
with the \textquotedblleft anagram\textquotedblright\ condition $%
\theta_{k}=\theta _{N+1-k}$, $(k=1,2,...,n)$. The composite Jones matrix of
the above described arrangement of wave plates in the LR basis is given by,
\begin{equation}
\mathbf{J}^{\left( N\right) }=\mathbf{J}_{\theta _{N}}\left( \varphi \right)
\mathbf{J}_{\theta _{N-1}}\left( \varphi \right) \cdots \mathbf{J}_{\theta
_{1}}\left( \varphi \right) .  \label{Jn}
\end{equation}%
Our objective is to implement an ideal half-wave plate propagator with Jones
matrix $\mathbf{J}_{0}$ in the LR basis (up to a global phase factor),
\begin{equation}
\mathbf{J}_{0}=\left[
\begin{array}{cc}
0 & i \\
i & 0%
\end{array}%
\right] ,
\end{equation}

\begin{table}[b]
\begin{tabular}{|c|l|}
\hline
$N$ & Rotation angles $(\theta _{1}$;\; $\theta _{2}$;\; $\cdots $;\; $%
\theta _{N-1}$;\; $\theta _{N}$) \\ \hline
3 & (60;\;120;\;60) \\ \hline
5 & (51.0;\;79.7;\;147.3;\;79.7;\;51.0) \\ \hline
7 & (68.0;\;16.6;\;98.4;\;119.8;\;98.4;\;16.6;\;68.0) \\ \hline
9 & (99.4;\;25.1;\;64.7;\;141.0;\;93.8;\;141.0;\;64.7;\;25.1;\;99.4) \\
\hline
11 & (31.2;\;144.9;\;107.8;\;4.4;\;44.7;\;158.6;\; \\
& 44.7;\;4.4;\;107.8;\;144.9;\;31.2) \\ \hline
13 & (59.6;\;4.9;\;82.5;\;82.7;\;42.7;\;125.8;\;147.2 \\
& 125.8;\;42.7;\;82.7;\;82.5;\;4.9;\;59.6) \\ \hline
\end{tabular}%
\caption{Rotation angles $\protect\theta _{k}$ (in degrees) for composite
broadband half-wave plates with different number $N$ of constituent
half-wave plates. }
\label{Table1}
\end{table}

by the product of half-wave plates $\mathbf{J}^{\left( N\right) }$ from Eq. %
\eqref{Jn}. That is, we set $\mathbf{J}^{(N)} \equiv \mathbf{J}_{0}$ at $%
\varphi =\pi $ which leaves us with $n$ independent angles $\theta _{k}$ to
use as control parameters. We then nullify as many lowest order derivatives
of $\mathbf{J}^{(N)}$ vs the phase shift $\varphi $ at $\varphi =\pi $ as
possible. We thus obtain a system of nonlinear algebraic equations for the
rotation angles $\theta _{k}$:
\begin{subequations}
\label{nullify}
\begin{eqnarray}
\left[ \partial _{\varphi }^{k}\mathbf{J}_{11}^{\left( N\right) }\right]
_{\varphi =\pi } &=&0\quad \left( k=1,2,...,n\right) ,  \label{Ja} \\
\left[ \partial _{\varphi }^{k}\mathbf{J}_{12}^{\left( N\right) }\right]
_{\varphi =\pi } &=&0\quad \left( k=1,2,...,n\right) .  \label{Jb}
\end{eqnarray}
\end{subequations}
The anagram symmetry assumption for the angles $\theta _{k}$ ($\theta
_{k}=\theta_{N+1-k}$), ensures that all even-order derivatives of $\mathbf{J}%
_{11}^{\left( N\right) }$ \eqref{Ja} and all odd-order derivatives of $%
\mathbf{J}_{12}^{\left( N\right) }$ \eqref{Jb} vanish;\; hence, the $n$
angles allow us to nullify the first $n$ derivatives of the matrix $\mathbf{J%
}^{(N)} $ \eqref{Jn}.

Solutions to Eqs. (\ref{nullify}) provide the recipe to construct \textit{%
arbitrary broadband } composite half-wave plates. A larger number $N$ of
ordinary half-wave plates provides a higher order of robustness against
variations in the phase shift $\varphi $ and thus, the light wavelength $%
\lambda $. We list several examples of broadband half-wave plates in Table %
\ref{Table1}.

\begin{figure}[th]
\centerline{\includegraphics[width=0.9\columnwidth]{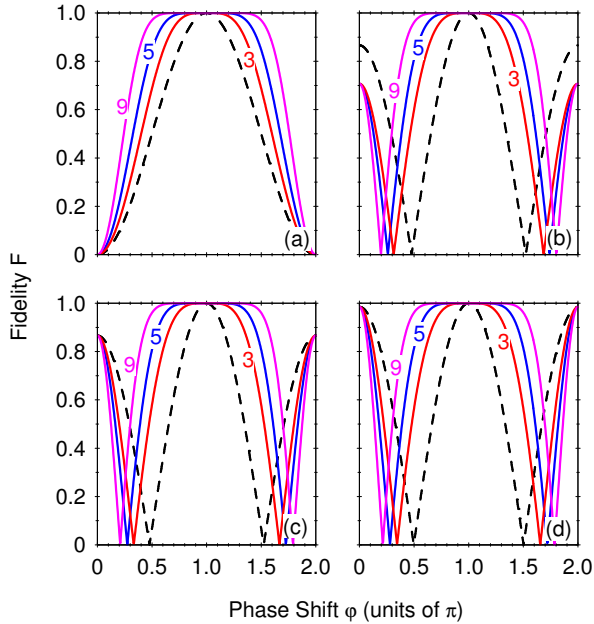}}
\caption{(Color online) Fidelity $F$ vs phase shift $\protect\varphi $ for
broadband polarization rotator, for different number of constituent plates $%
N $ in the composite half-wave plates. The rotation angles of the two
composite half-wave plates are those given in Table \protect\ref{Table2}.
Frame (a) for polarization rotation of $90$ degree, frame (b) for
polarization rotation of $45$ degree, frame (c) for polarization rotation of
30 degree, and frame (d) for polarization rotation of 10 degree. The black dashed line is for a rotator composed of just two half-wave plates, given for easy reference.}
\label{Fig2}
\end{figure}

\section{Broadband rotator}


We proceed to show how a broadband rotator can be constructed as a sequence
of two broadband half-wave plates with an additional rotation between them.
Let us first consider a simple sequence of two ordinary half-wave plates
with a relative rotation angle $\alpha /2$ between them. We multiply the
Jones matrices of the two half-wave plates ($\varphi =\pi $) given in Eq. (%
\ref{retarder}), where one is rotated at $\alpha /4$ while the other is
rotated at $-\alpha/4 $. Thus, we obtain the total propagator
\begin{equation}
\mathbf{J}_{\alpha /4}(\pi )\mathbf{J}_{-\alpha /4}(\pi )=-\left[
\begin{array}{cc}
e^{i\alpha } & 0 \\
0 & e^{-i\alpha }%
\end{array}%
\right] ,  \label{rotator2}
\end{equation}%

which represents a Jones matrix for a rotator in the LR basis (up to
unimportant $\pi$ phase),
\begin{equation}
\mathbf{R}(\alpha )=\left[
\begin{array}{cc}
e^{i\alpha } & 0 \\
0 & e^{-i\alpha }%
\end{array}%
\right] .  \label{rotator Jones matrix}
\end{equation}

However, a rotator constructed from two ordinary half-wave plates according
to Eq. \eqref{rotator2} is not broadband. We overcome this limitation and
extend the range of operation of the rotator by replacing the two ordinary
half-wave plates with two identical broadband composite half-wave plates as
described in Section \ref{sec3}. In order to calculate the efficiency of our
composite polarization rotator we define the fidelity $F$ \cite%
{Ivanov,Ardavan} according to,
\begin{equation}
F=\frac{1}{2}\text{Tr}\left( \mathbf{R}^{-1}(\alpha )\mathbf{J}_{\mathnormal{%
tot}}\right),
\end{equation}%
where $\mathbf{J}_{\mathnormal{tot}} = \mathbf{J}_{\alpha /4}^{(N)}\mathbf{J}%
_{-\alpha/4}^{(N)}\mathbf{\ }$ with $\mathbf{J}_{\alpha /4}^{(N)}$ and $%
\mathbf{J}_{-\alpha /4}^{(N)}$ representing the $N$-composite broadband
half-wave plates from Eq. \eqref{Jn} rotated additionally to $\alpha /4$ and
$-\alpha /4$, respectively.

\begin{figure}[h]
\includegraphics[width=0.9\columnwidth]{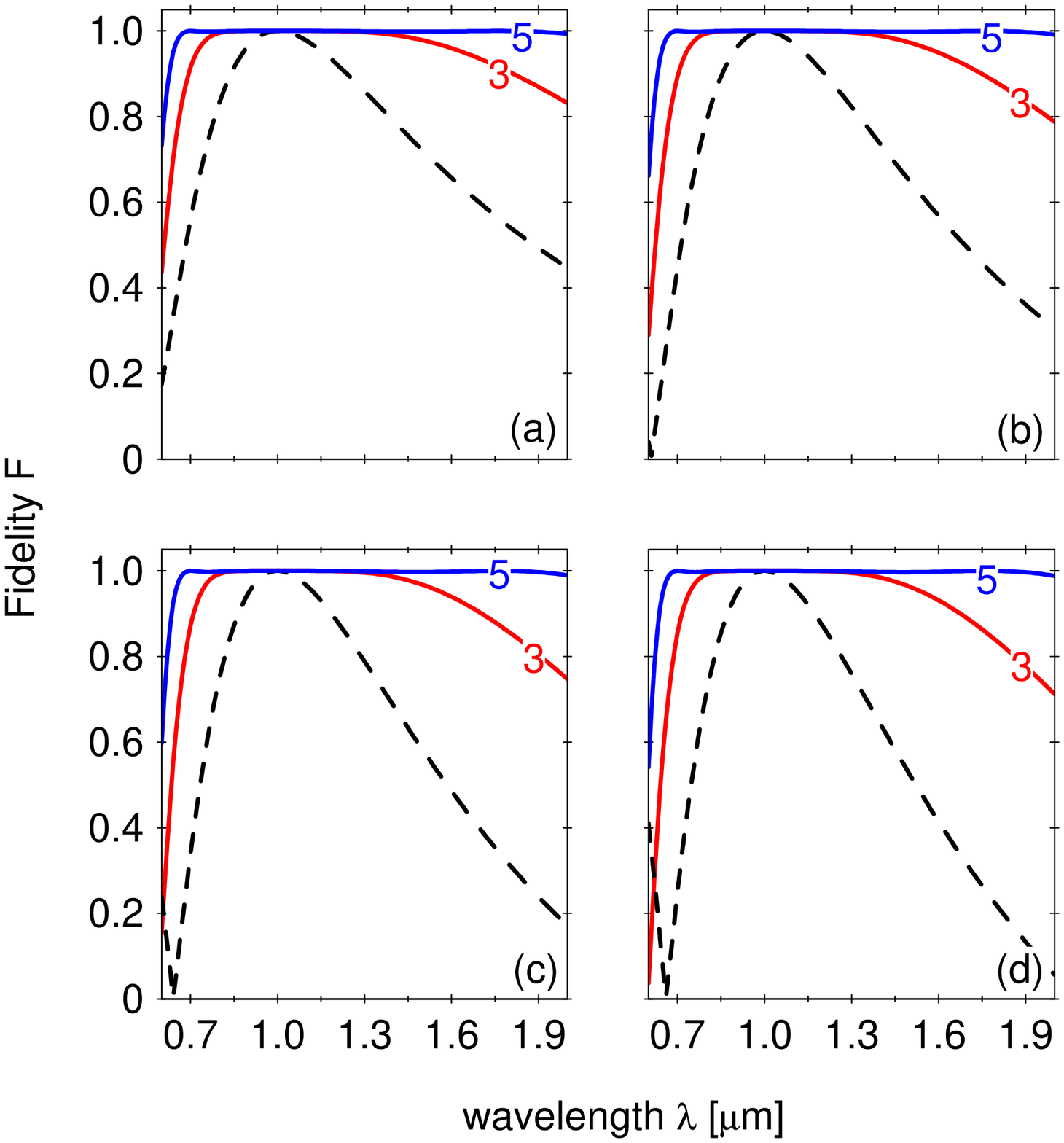}
\caption{(Color online) Fidelity $F$ vs wavelength $\lambda $ for
broadband polarization rotator, for different number of constituent plates $%
N $ in the composite half-wave plates. The rotation angles of the two
composite half-wave plates are those given in Table \protect\ref{Table2}.
Frame (a) for polarization rotation of $90$ degree, frame (b) for
polarization rotation of $75$ degree, frame (c) for polarization rotation of
60 degree, and frame (d) for polarization rotation of 45 degree. The black dashed line is for a rotator composed of just two half-wave plates, given for easy reference.}
\label{Fig3}
\end{figure}
\begin{table*}[hbt]
\begin{tabular}{|c|l|}
\hline
$N+N$ & Rotation angles of composite half-wave plates ($\theta _{1}$-$\alpha
/4;$ $\theta _{2}$-$\alpha $/4; $\cdots $;\;$\theta _{N}$-$\alpha $/4)$%
(\theta _{1}$+$\alpha $/4;\;$\theta _{2}$+$\alpha $/4;\;$\cdots $;\;$\theta
_{N}$+$\alpha $/4) \\ \hline
3+3 & (60-$\alpha $/4;\;120-$\alpha $/4;\;60-$\alpha $/4)(60+$\alpha $%
/4;\;120+$\alpha $/4;\;60+$\alpha $/4) \\ \hline
5+5 & (51.0-$\alpha $/4;\;79.7-$\alpha $/4;\;147.3-$\alpha $/4;\;79.7-$%
\alpha $/4;\;51.0-$\alpha $/4)(51.0+$\alpha $/4;\;79.7+$\alpha $/4;\;147.3+$%
\alpha $/4;\;79.7+$\alpha $/4;\;51.0+$\alpha $/4) \\ \hline
7+7 & (68.0-$\alpha $/4;\;16.6-$\alpha $/4;\;98.4-$\alpha $/4;\;119.8-$%
\alpha $/4;\;98.4-$\alpha $/4;\;16.6-$\alpha $/4;\;68.0-$\alpha $/4) \\
& (68.0+$\alpha $/4;\;16.6+$\alpha $/4;\;98.4+$\alpha $/4;\;119.8+$\alpha $%
/4;\;98.4+$\alpha $/4;\;16.6+$\alpha $/4;\;68.0+$\alpha $/4) \\ \hline
9+9 & (99.4-$\alpha $/4;\;25.1-$\alpha $/4;\;64.7-$\alpha $/4;\;141.0-$%
\alpha $/4;\;93.8-$\alpha $/4;\;141.0-$\alpha $/4;\;64.7-$\alpha $/4;\;25.1-$%
\alpha $/4;\;99.4-$\alpha $/4) \\
& (99.4+$\alpha $/4;\;25.1+$\alpha $/4;\;64.7+$\alpha $/4;\;141.0+$\alpha $%
/4;\;93.8+$\alpha $/4;\;141.0+$\alpha $/4;\;64.7+$\alpha $/4;\;25.1+$\alpha $%
/4;\;99.4+$\alpha $/4) \\ \hline
\end{tabular}%
\caption{Rotation angles $\protect\theta _{k}$ (in degrees) for each
half-wave plate in the composite broadband polarization rotator for
arbitrary rotation angle $\protect\alpha $. }
\label{Table2}
\end{table*}
Several examples of broadband polarization rotators, using the composite
broadband half-wave plates from Table \ref{Table1}, are given in Table \ref%
{Table2}. Their fidelity with respect to variations in the phase shift $%
\varphi $ is illustrated in Fig.\ref{Fig2}.

Furthermore, in Fig. \ref{Fig3} we illustrate
the fidelity of the suggested broadband polarization
rotator, using true zero order half-wave plates, which operate around 1 $\mu m$, made
from quartz (thickness $L$ of each wave plate is 57 $\mu m$). The retardance
of the wave plates is calculated using the Sellmeier equations for ordinary
and extraordinary refractive indexes \cite{Ghoshr}.


\section{Conclusion}


We have presented an approach to construct an arbitrary broadband composite
polarization rotator which can rotate the polarization of a linearly polarized light at any desired angle $\alpha$. The polarization rotator is comprised of two composite broadband half-wave plates with a relative rotation of $\alpha /2$ between them. Furthermore, using a sequence of zero order quartz half-wave plates, we numerically showed that the polarization rotator is broadband and operates over a wide range of wavelengths. An experimental implementation of the suggested broadband polarization rotator with half-wave plates which are
readily available in most laboratories should be straightforward. Finally,
we note that this technique has an analogue in atomic physics in terms of
composite phase gates as recently was demonstrated by Torosov and Vitanov
\cite{Torosov}.

\acknowledgments

We acknowledge financial support by Singapore University of Technology and Design Start-Up Research Grant, Project no. SRG-EPD-2012-029
and SUTD-MIT International Design Centre (IDC) Grant, Project no. IDG31300102. The
authors are grateful to Svetoslav Ivanov for useful discussions.


\end{document}